\documentclass[prl,twocolumn,amsmath,amssymb,floatfix]{revtex4}

\usepackage[utf8]{inputenc}

\usepackage{amsmath}
\usepackage{graphicx}
\usepackage{bm}

\newcommand{\be}{\begin{equation}}
\newcommand{\ee}{\end{equation}}
\newcommand{\la}{\langle \!\langle}
\newcommand{\ra}{\rangle \!\rangle}

\newcommand{\beq}{\begin{eqnarray}}
\newcommand{\eeq}{\end{eqnarray}}

\def\eq#1{(\ref{#1})}
\def\H1{\widehat{H}_1}

\def\op#1{{\Hat{\mathrm{#1}}}}
\def\vop#1{{\Hat{\mathbf{#1}}}}

\def\ket#1{\ensuremath{\vert{#1}\rangle}}

\begin{document}

\title{Supersymmetry in quantum optics and in spin-orbit coupled systems}

\author{Michael Tomka$^{1}$, Mikhail Pletyukhov$^2$, Vladimir Gritsev$^3$}
\affiliation{$^1$Physics Department, University of Fribourg, Chemin du
Mus\'{e}e 3, 1700 Fribourg, Switzerland\\
$^2$Institute for Theory of Statistical Physics and JARA -- Fundamentals of Future Information Technology, 
RWTH Aachen, 52056 Aachen, Germany \\
$^3$ Institute of Theoretical Physics, University of Amsterdam, Science park 904, 1098 XH Amsterdam, Netherlands}

\begin{abstract}
Light-matter interaction is naturally described by coupled bosonic and
fermionic subsystems.
This suggests that a certain Bose-Fermi duality is naturally
present in the fundamental quantum mechanical description of photons
interacting with atoms.
We reveal submanifolds in parameter space of a basic light-matter interacting system where this duality is promoted to  a
supersymmetry (SUSY) which remains unbroken.
We show that SUSY is robust with respect to decoherence and
dissipation.
In particular, a stationary density matrix at the supersymmetric lines in
the parameter space has a degenerate subspace.
A dimension of this subspace is given by the Witten index and thus
topologically protected.
As a consequence of this SUSY, dissipative dynamics at the supersymmetric
lines is constrained by an additional conserved
quantity which translates some part of 
information about an initial state into the stationary state
subspace. 
We also demonstrate a robustness of this additional conserved quantity away from
the supersymmetric lines.
In addition, we demonstrate that the same SUSY structures are present
in condensed matter systems with spin-orbit couplings of Rashba and
Dresselhaus types, and therefore spin-orbit coupled systems at the
SUSY lines should be robust with respect to various types of disorder
and decoherences.
Our findings suggest that optical and condensed matter systems at the
SUSY points can be used for quantum information technology and can
open an avenue for quantum simulation of the SUSY field theories.
\end{abstract}

\maketitle


{\it Introduction.---}A concept of supersymmetry (SUSY) is one of
the most beautiful and attractive in physics, since it establishes a
duality between bosons and fermions, cures divergency problems and
resolves the mass hierarchy in quantum field theory~\cite{weinberg}. 
Furthermore, in cosmology it can serve as an explanation of the
dark matter essence~\cite{susy-cosmology}.
It exists in nature, this symmetry must be broken, since there is no so far known phenomenon in which a boson is converted into a fermion. 
Therefore, to observe its signatures it is believed that we
need powerful accelerators.
However, recent progress with quantum simulators using synthetic
matter (like e.g.\ cold atoms, ion traps and coupled cavities systems)
allows us to think in the direction of realizing supersymmetric
systems in the nowadays laboratory. 
Here we show that SUSY systems can be engineered in simple
and fundamental models either by means of solid state devices or
by quantum optical schemes.
One implementation we discuss is based on a generalized version of the
Rabi model of quantum optics, while the other one is based on the 2D
electron gas in a magnetic filed with the Rashba and Dresselhaus
spin-orbit coupling. 
Further, we reveal that the manifolds in parameter space where the
SUSY is unbroken are robust with respect to dissipation and
decoherence. 
This suggest that SUSY systems have an advantage for being used in
quantum information science.

\begin{figure}
\begin{center}
\includegraphics[width=\columnwidth]{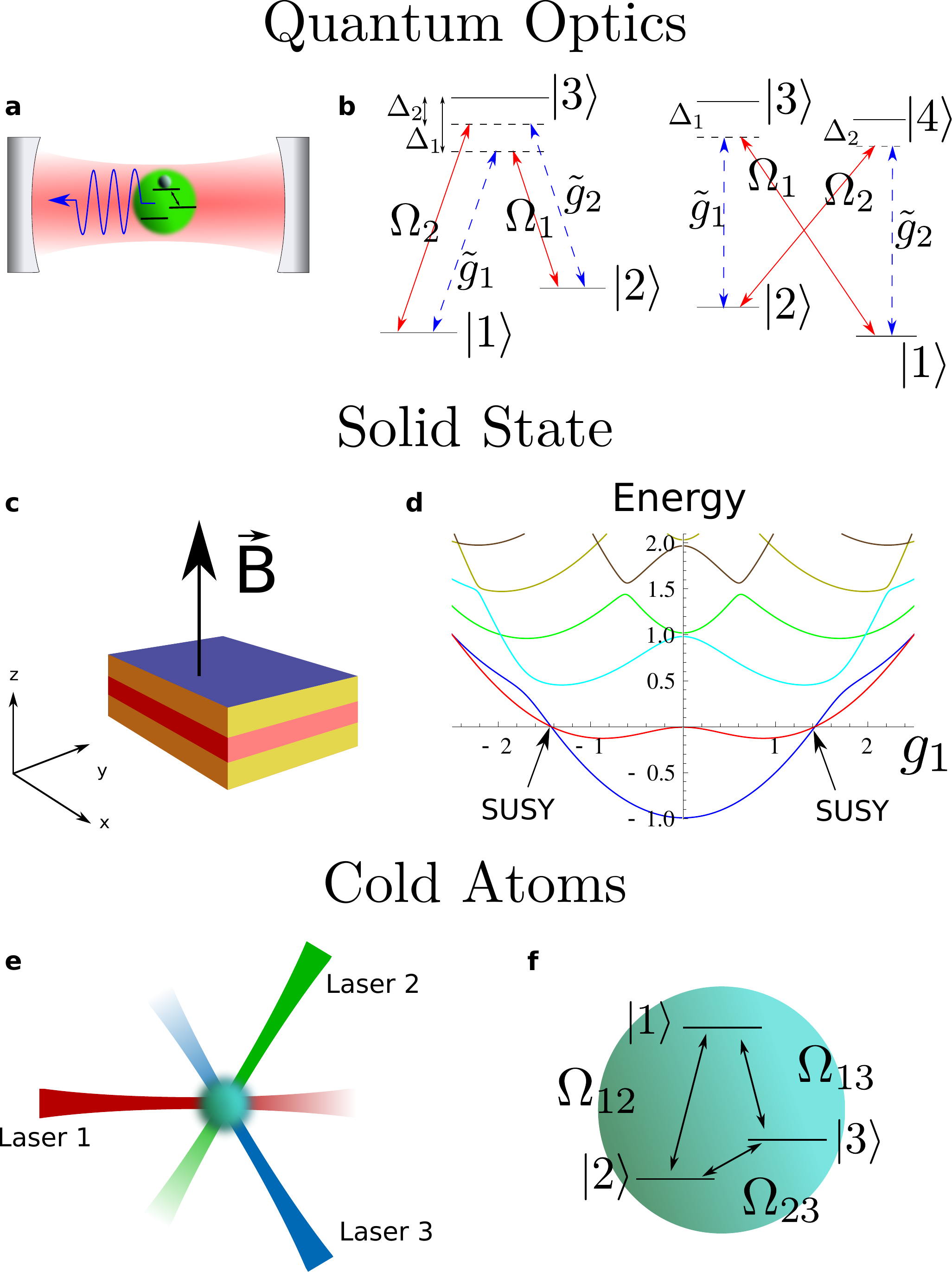}
\caption{In the field of quantum optics SUSY appears in a generalized
  Rabi model which can be realized in cavity-QED systems \textbf{a}
  using the $\Lambda$-type 3- or 4-level transition schemes
  \textbf{b}.
  In solid state systems the 2D electron gas with Rashba and Dresselhaus spin-orbit
  couplings subject to a perpendicular magnetic field \textbf{c} can
  also be mapped to the Rabi model with unequal couplings of the co-
  and counter-rotating terms.
  In \textbf{d} we show the energy spectrum of these models as a function
  of the coupling parameter $g_{1}\sim\alpha_{R}$ and for
  $\alpha_{D} \sim g_{2}=0.2$, the SUSY lines occur when the
  parameters $g_{1}^{2}-g_{2}^{2}=\Delta\omega$, in terms of Eq.~(\ref{gR}).
  In Ref.~\cite{socreview} a possible realization of
  tunable Rashba and Dresselhaus SOC with ultracold alkali atoms is
  proposed \textbf{e}, where each state is coupled by a two-photon
  Raman transition \textbf{f}.}
  \label{fig:illustration}
\end{center}
\end{figure}

The role of spin-orbit (SO) coupling is central for a number of
current developments in low-dimensional materials: spin Hall effect,
anomalous Hall effect, spintronics~\cite{spintronics}, topological
insulators and superconductors~\cite{top-ins} and Majorana
fermions~\cite{majorana}.
Recently, synthetic gauge fields and SO coupling has also been
realized in ultracold Bose and Fermi gases with Raman
beams~\cite{so-cold-gases}. 
Behind all these developments stands a simple single-particle model
which is identified by the names of Rashba and Dresselhaus.
We found here that spin-orbit coupled systems can have SUSY in a broad
range of parameters.

In the field of quantum optics an even more fundamental role is played
by the Jaynes-Cummings and Rabi models. 
These models describe a system of a single bosonic mode coupled to a
two-level system via dipole interaction.
The understanding of the dynamics in these models led to a
breakthrough in cavity-QED systems~\cite{cav-QED-1}, nano photonics,
etc. 
The whole presence of bosons (light quanta) and fermions (two-level
systems) suggests that there is a hidden supersymmetry in the quantum
optical models.

In this paper we reveal explicitly a SUSY structure present in a generalized
version of the Rabi model of quantum optics.
Further, we show that the generalized Rabi model can be realized in a
two dimensional electron gas with Rashba and Dresslhaus spin-orbit
coupling subject to a perpendicular and constant magnetic field.
The influence of this SUSY on the dissipative dynamics of the
generalized Rabi model is studied.
We observed that due to the supersymmetry the dissipative dynamics,
governed by the master equation in the dressed state picture, shows an
additional conserved quantity when the system is supersymmetric.
Further, we studied the behavior of this additional conserved quantity
if the system slightly deviates from the supersymmetry.

{\it Model and its realizations.---}We consider one of the simplest
and most fundamental model describing the interaction of a single mode
bosonic field (represented by the canonical operators $\op{a},
\op{a}^{\dag}$) with a single two-level system (described by the Pauli
matrices $\op{\sigma}_{i}$, $i=\pm,z$),
\beq
\op{H}_{\mathrm{gR}} &=&
\hbar \, \omega \, \op{a}^{\dag}\op{a}
+
\frac{\Delta}{2} \op{\sigma}_{z} \nonumber \\
&+&
g_{1} ( \op{a}^{\dag}\op{\sigma}_{-} + \op{a} \op{\sigma}_{+} )
+
g_{2} ( \op{a}^{\dag}\op{\sigma}_{+} + \op{a} \op{\sigma}_{-} ).
\label{gR}
\eeq
The energies of the bosonic field and the energy splitting of the
two-level system are $\omega$ and $\Delta$ respectively, while the
interaction constants $g_{1,2}$ can be arbitrary and real.

In the realm of quantum optics the model~(\ref{gR}) describes a
single mode electromagnetic field interacting with a two-level emitter
via dipole interaction and represents a direct generalization of two
fundamental models in quantum optics. 
Namely, when either $g_{2}=0$ or $g_{1}=0$ it is known as the Jaynes-Cummings
model~\cite{JC} while when $g_{1}=g_{2}$ it becomes the Rabi
model~\cite{Rabi}.
In these limits a number of spectral and dynamical properties are
known while it is much less studied for arbitrary $g_{1,2}$.
In the weak coupling regime close to resonance $\omega \sim \Delta$
only the $g_{1}$ term is relevant and $g_{2}$ scales to zero
(RWA, Rotating Wave Approximation). 
On the contrary, when the strong coupling regime is realized, both co-
and counter-rotating terms have to be kept.
When the Rabi model is derived from the microscopic principles then
the coupling constants are such that $g_{1}=g_{2}$.
The Jaynes-Cummings model was studied extensively in the literature
and can be solved exactly since the total number of excitations
$\op{N}=\op{a}^{\dag}\op{a}+\op{\sigma}_{z}/2$ is a conserved quantity. 
In contrast, the analytical solution of the Rabi model is still under
active discussions~\cite{braak}, despite the long history of the
model. Similarly to the Rabi model, the Hamiltonian $(\ref{gR})$ commutes with the parity operator $P=\exp(i\pi \op{N})$. 
While the spectrum of the Jaynes-Cummings model ($g_{2}=0$) is well known
the spectrum of the Rabi model ($g_{1}=g_{2}$) is given by a
self-consistent set of equations which can be solved
numerically~\cite{braak}. 
We note that in the limit of strong coupling (both $g_{1,2}/\omega$ are large) the spectrum consist of a two quasi-degenerate harmonic ladders~\cite{irish}.
%
Both models are of immense experimental interest for
cavity-~\cite{cav-QED-1} and circuit-QED setups, superconducting q-bits, NV
centers, etc. 
The solid-state devices are able to approach a strong-coupling regime
where $g_{2}$ term becomes relevant \cite{circ-QED-2},\cite{circ-QED-3}.
In in the field of quantum optics the model with unequal $g_{1}$ and
$g_{2}$ can be realized using the $\Lambda$-type 3- or 4-level
transition schemes~\cite{3-level},~\cite{4-level}, see also
Fig.~\ref{fig:illustration}.

Indeed, considering two non-degenerate ground states $\ket{1}$,
$\ket{2}$ coupled by a quantum field $\op{a}$ with couplings
$\tilde{g}_{1,2}$ to the excited state(s) and, in addition,
introducing two classical Rabi couplings $\Omega_{1,2}$ one can
perform an adiabatic elimination of the excited states.
The resulting effective Hamiltonian then has the form of the
generalized Rabi model ($=\op{H}_{\mathrm{gR}}$) with an additional
Bloch-Siegert shift, $\op{H}_{\mathrm{eff}} = \op{H}_{\mathrm{gR}} +
\lambda \, \op{a}^{\dag}\op{a} \, \op{\sigma}_{z}$
(see the Supplement for an overview of the  derivation).
%
%
The parameters of the generalized Rabi model are then given by
$g_{1,2}=\tilde{g}_{1,2}\Omega^{*}_{1,2}/2\Delta_{1,2}$,
$\omega=(\tilde{g}_{1}^{2}/\Delta_{1}+\tilde{g}_{2}^{2}/\Delta_{2})/2$,
$\Delta=(\Omega^{2}_{1}/\Delta_{1}-\Omega^{2}_{2}/\Delta_{2})$ and
$\lambda=g_{1}^{2}/\Delta_{1}-g_{2}^{2}/\Delta_{2}$. 
While the couplings $\tilde{g}_{1,2}$ are predefined, the Rabi
frequencies $\Omega_{1,2}$ as well as the detunings $\Delta_{1,2}$ can
be tuned in a wide range and therefore we can consider the
model~(\ref{gR}) for variable $g_{1,2}$.
If we choose the parameters to cancel the Bloch-Seigert shift,
$\frac{g_{1}^{2}}{\Delta_{1}}=\frac{g_{2}^{2}}{\Delta_{2}}$ we end up
with the generalized Rabi model~(\ref{gR}).
In Ref.~\cite{grimsmo4level} a simulation of the Rabi model with
unequal $g_{1}$ and $g_{2}$ and with an effective Bloch-Siegert shift
($\propto \op{a}^{\dag}\op{a} \, \op{\sigma}_{z}$) is proposed based on
the resonant Raman transitions in an atom that interacts with a high
finesse optical cavity mode (four-level transition scheme).

The same model appears in various branches of condensed matter science
where the spin-orbit coupling plays an import an role.
In particular, this is the case for a two-dimensional noninteracting electron
system with Rashba and Dresselhaus spin-orbit coupling in a perpendicular magnetic field.
In solid state devices this can be realized either by the electron gas
in the quantum wells, in 2D topological insulators or in the quantum
dots with parabolic confinement potential. 
In cold atoms spin-orbit coupling can be achieved
artificially~\cite{so-cold-gases},~\cite{socreview}.
For the case of a two-dimensional electron gas subject to a
perpendicular magnetic field ${\bf B}=B_{0}{\bf e}_{z}$, the spin-orbit
coupled Hamiltonian reads
\begin{align}
\op{H}_{\mathrm{RD}}
&=
\frac{\op{\Pi}^{2}_{x} + \op{\Pi}^{2}_{y}}{2m^{*}} + g^{*}\mu_{B}B_{0}\op{\sigma}_{z} \nonumber \\
&+
\frac{2\alpha_{R}}{\hbar}(\op{\Pi}_{x}\op{\sigma}_{y} - \op{\Pi}_{y}\op{\sigma}_{x})
+
\frac{2\alpha_{D}}{\hbar}(\op{\Pi}_{x}\op{\sigma}_{x} + \op{\Pi}_{y}\op{\sigma}_{y}),
\label{R+D}
\end{align}
where $\op{\Pi}_{x}=\op{p}_{x}-\frac{eB_{0}}{2c}\op{y}$,
$\op{\Pi}_{y}=\op{p}_{y}+\frac{eB_{0}}{2c}\op{x}$ are momentum
operators in symmetric gauge, $\alpha_{R}$ represents the Rashba
spin-orbit coupling, while $\alpha_{D}$ denotes the Dresselhaus
spin-orbit coupling, $m^{*}$ is the effective electron mass, $g^{*}$
is the gyromagnetic ratio and $\mu_{B}=e\hbar/2m$ is the Bohr
magneton. 
A short derivation of the connection between (\ref{R+D}) and
(\ref{gR}) is reproduced in the Supplement.
%
%
In this way we established an equivalence between the electronic
Rashba+Dresselhaus model with a magnetic field and the
Jaynes-Cummings-Rabi, which we called the generalized Rabi model (gR
in the following), from quantum optics. 

The correspondence $\op{H}_{\mathrm{RD}}=\op{H}_{\mathrm{gR}}$ has a
potential to cross-fertilize the two areas of research, where these
models play a fundamental role: those branches of condensed matter
physics where the spin-orbit coupling plays a crucial role and the
field of quantum optics.
Some examples of this we show below by looking at the quench dynamics
in both models.




{\it Supersymmetry.---} Supersymmetric filed theories which were
studied intensively during the last 40 years have a supersymmetric
quantum mechanics (SUSY QM) as their low-energy limit.
Introduced in 70's the SUSY quantum mechanics became a subfield by itself~\cite{Junker,CKS} with many applications.
Here we are interested in the $N=2$ SUSY QM. 
This SUSY QM is characterized by two supercharges $\op{Q}_{1}$ and
$\op{Q}_{2}$ that satisfy the algebra $\{\op{Q}_{i},\op{Q}_{j}\}=2\delta_{ij}\vop{H}$ ($i,j=1,2$), where $\vop{H}$
is known as the supersymmetric Hamiltonian.
One can also introduce the linear combinations of the supercharges
$\op{Q} = \op{Q}_{1} + i \op{Q}_{2}$, $\op{Q}^{\dag}=\op{Q}_{1}-i\op{Q}_{2}$ such that
$\op{Q}^{2}=(\op{Q}^{\dag})^{2}=0$ and $\{\op{Q}, \op{Q}^{\dag}\} = \vop{H}\equiv\mbox{diag}(H_{+},H_{-})$. 
The Witten parity operator
$\op{W}=[\op{Q},\op{Q}^{\dag}]/\{\op{Q},\op{Q}^{\dag}\}$ commutes
with the SUSY Hamiltonian, anti-commutes with the supercharges and has
the eigenvalues $\pm 1$. It distinguish between two super partner Hamiltonians $H_{\pm}$ which have the same energy spectrum except for the ground state.
The supersymmetry is called \textit{unbroken} if all the (degenerate) ground states are annihilated
$\op{Q}\ket{\psi_{0}^{j}} = 0$, $\forall j$, whereas if the SUSY is
\textit{broken} then there exists at least one quantum state for which
$E_{0} > 0$. Usually only one of $H_{\pm}$ has zero modes if SUSY is unbroken.

Our findings can be summarized as follows: (1) SUSY as a {\it symmetry}
exists in the gR model for a special combination of parameters,
\be
g_{1}^{2}-g_{2}^{2}=\Delta\omega
\label{susy-line}
\ee
when the Bloch-Siegert shift is zero, $\lambda=0$, (in the the special case of
$g_{1}=g_{2}$ SUSY exists for degenerate atomic levels, $\Delta=0$,
and in this case the Hamiltonian has the form of a shifted harmonic oscillator). 
The associated supercharges in matrix representation are given by
\be
\op{Q}
=
\begin{pmatrix}
 0 & \op{q}  \\
 0 & 0
\end{pmatrix},
\qquad
\op{q}
=
\begin{pmatrix}
 \frac{g_{1}}{\sqrt{\omega}} & \sqrt{\omega} \, \op{a} \\
 \sqrt{\omega} \, \op{a}  & \frac{g_{2}}{\sqrt{\omega}}
\end{pmatrix}.
\ee
At this line SUSY Hamiltonians $H_{\pm}=H_{gR}+const$, as demonstrated in the Supplement.
When $\lambda\neq 0$ the SUSY condition reads $2(\Delta-\lambda)(\omega-\lambda)(\omega+\lambda)=g_{1}^{2}(\omega-\lambda)-g_{2}^{2}(\omega+\lambda)$.
(2) On the line (\ref{susy-line}) in the parameter space the SUSY is
unbroken and the Hamiltonian has a doubly-degenerate ground state. This
implies that the Witten index is equal two. 
The Witten index $W_{\mathrm{ind}}$ is given by the difference
between the zero eigenmodes $n_{\pm}$ of $\op{H}_{\pm}$, namely
$W_{\mathrm{ind}} = \dim \ker \op{H}_{-} - \dim \ker \op{H}_{+} = n_{-} - n_{+}$.
It is related to the index of the annihilation operator $\op{q}$, i.e.
$W_{\mathrm{ind}} = \mathrm{ind} \, \op{q} = \dim \ker \op{q} - \dim \ker \op{q}^{\dag}$,
and has a property of {\it topological} invariance~\cite{atiyah} according to the Atiyah-Singer index theorem.
%
%
We show explicitly in the Supplement that there are two zero
eigenmodes of $\op{H}_{-}$, and zero for $\op{H}_{+}$, thus
$W_{\mathrm{ind}}=2$.  Similarly to the Rabi case, the Hamiltonian (\ref{gR}) commutes with the parity operator $P$; therefore two zero modes are the eigenstates of the parity operator $P$
and can be written as $|\psi_{0}\rangle =[D(\alpha)\pm D(-\alpha)]|0\rangle$ where $D(\alpha)$ is a coherent state displacement operator with $\alpha=g_{1}g_{2}/\sqrt{\omega}$.
The explicit derivation of the supercharges and zero modes for the gR model are
given in the Supplement.

{\it Dissipative dynamics.---}In the quantum optical realization of gR
model the effects of coupling the system to the environment are usually accounted
for by the master equation in the Lindblad form.
Here we show that the SUSY in the gR model is stable against couplings to several
types of dissipative baths.
Effects of relaxation and decoherence are described by the Lindblad master
equation for the density matrix in the dressed
picture~\cite{CW},~\cite{BP},~\cite{BGB},~\cite{AHE}:
$\partial_{t}\op{\rho}=-i[\op{H}_{\mathrm{gR}},\op{\rho}]+{\cal L}_{dr}\op{\rho}$ where the dissipator
${\cal L}_{dr}$ should be written in terms of the jump operators
$| j \rangle \langle k|$ between the exact eigenstates $| j \rangle$,
$| k \rangle$ of the Hamiltonian,
$\op{H}_{\mathrm{gR}} | j \rangle = \epsilon_{j} | j \rangle$,
\begin{align}
{\cal L}_{dr}
&=
{\cal D}[\sum_{j}\Phi^{j}|j\rangle\langle j|]
+
\sum_{j,k:k>j}(\Gamma^{jk}_{\kappa}+\Gamma^{jk}_{\gamma}){\cal D}[|j\rangle\langle k|]
\label{diss}
\end{align}
where ${\cal D}[\op{O}]\op{\rho} = \frac{1}{2}(2
\op{O}\op{\rho}\op{O}^{\dag} - \op{\rho}\op{O}^{\dag}\op{O} -
\op{O}^{\dag}\op{O}\op{\rho})$ is a quantum dissipator.
The different terms in Eq.~(\ref{diss}) correspond to different
sources of decoherence:
The first term
$\Phi_{j}=\sqrt{\gamma_{\phi}(0)/2} \langle j |\op{\sigma}_{z}| j \rangle$ 
describes the diagonal part of the dephasing of the two-level system
in the eigenbasis and $\gamma_{\phi}(0)$ is the dephasing rate
quantified by the dephasing noise spectral density at zero frequency.
The other two terms describe contributions from the oscillator and the
two-level system baths.
They cause transitions between eigenstates with the relaxation
coefficients
%
%
%
%
$\Gamma^{jk}_{c} = 2 \pi d_{c}(\Delta_{kj})\alpha^{2}_{c}(\Delta_{kj})|C^{(c)}_{jk}|^{2}$, 
where $d_{c}(\Delta_{kj})$ is the spectral density of the bath and
$\alpha_{c}(\Delta_{kj})$ is the system-bath coupling strength at the
transition frequency $\Delta_{kj}=\epsilon_{k}-\epsilon_{j}$.
The transition coefficients are 
$C_{jk}^{(c)} = \langle j | \op{c} + \op{c}^{\dag} | k \rangle$ with $\op{c}=\op{a},\op{\sigma}_{-}$.
The spectral density can be assumed to be constant while
$\alpha^{2}_{c}(\Delta_{kj}) \propto \Delta_{kj}$.
Hence
$\Gamma^{jk}_{c}=\gamma_{c}\frac{\Delta_{kj}}{\omega}|C^{c}_{jk}|^{2}$,
where $\gamma_{c} \equiv \kappa_{c},\gamma_{c}$ are the standard
damping rates of a weak coupling scenario for the bosonic and spin
channels of dissipation~\cite{BGB}.

\begin{figure}[h!]
  \begin{center}
      \includegraphics[scale=1.0]{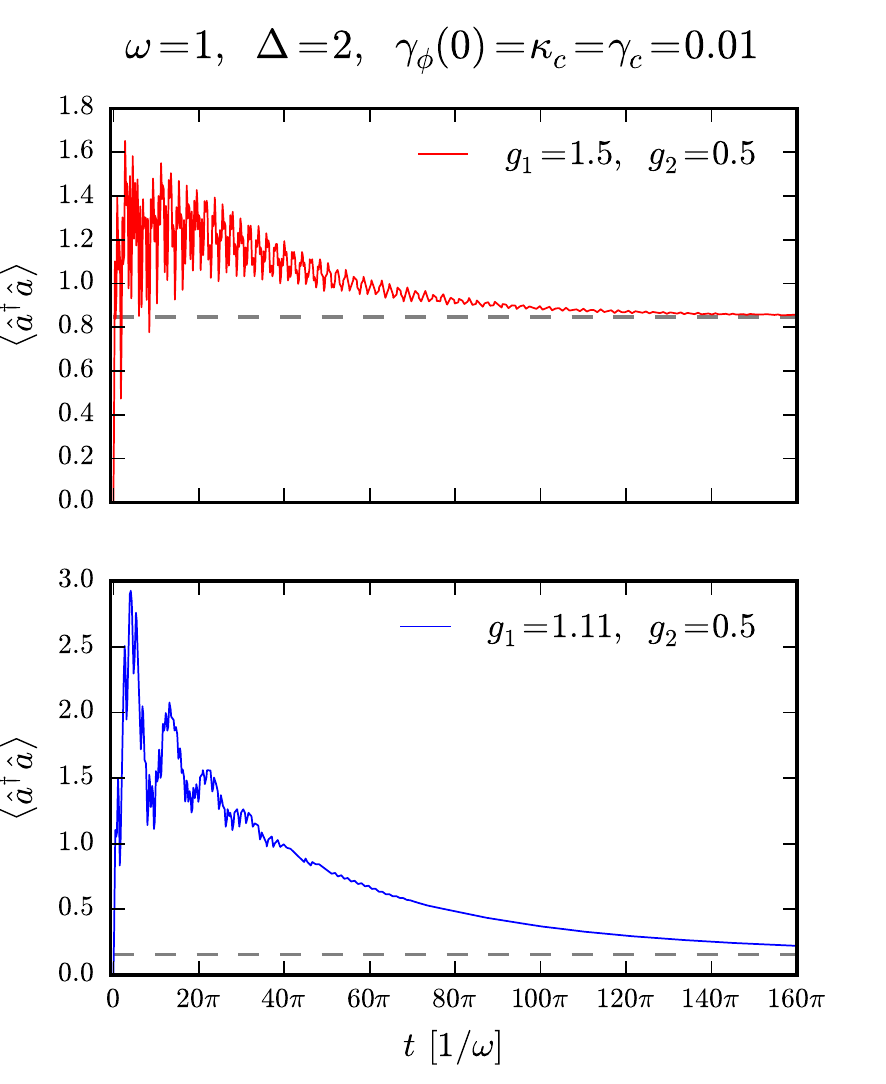}
    \caption{Dissipative dynamics of the generalized Rabi model:
             The time evolution of the mean-photon number for the initial state $|0\rangle_{b}|\uparrow\rangle$ (zero photons and excited two-level
             system).
             Upper panel: evolution for parameters of the model tuned to the SUSY line (\ref{susy-line}). The stationary value (dashed
             line) computed with the help of the conserved quantity
             $I_{1}$ and $I_{2}$.
             Lower panel: dissipation far away from the SUSY line.
             In this case the stationary state is given by the $I_{1}$
             which corresponds to the trace and gives the ground state
             expectation value.
             } 
    \label{fig:diss}
  \end{center}
\end{figure}
Using the dressed-picture dissipative formalism we checked that the
dynamics preserved the trace property and that the ground state
evolution has no time dependence.
In Fig.~\ref{fig:diss} we illustrate the time evolution of the mean-photon
number when the initial state is taken in the
``spin up'' state with zero bosonic occupation.
The evolution at the SUSY line exhibits oscillatory behavior, while
away from the SUSY line the dynamics is damped.

Usually a dissipative quantum system has a unique limit for the stationary state density matrix. However this is not always the case. 
Here we found that the stationary solution of the density matrix
equation has a {\it manifold of stationary states} at the SUSY line. Namely, the stationary solution of the Lindblad equation ${\cal L}_{dr}\hat{\rho}_{st}-i[\hat{H}_{gR},\hat{\rho}_{st}]=0$ has four-fold degenerate zero eigenvalue when $\gamma_{\phi}(0)=0$. This manifold of the the density matrix stationary states is spanned by the operators $|i\rangle\langle j|$, where $i,j=1,2$ label two degenerate states, and thus the manifold of the stationary states is equivalent to the space of unit quaternions, and can be parametrized by the $SU(2)$ group. On the other hand, when $\gamma_{\phi}(0)\neq 0$ only diagonal part of this $SU(2)$ matrix survives and the stationary state is only doubly degenerate. In Supplement we demonstrate that the dimension of space of the stationary density matrix is {\it topologically protected} for zero dephasing. 
As a consequence of the degenerate stationary subspace there is, in
addition to the trace, another {\it conserved quantity} commuting with the Liouvillian. These conserved quantities are constructed as an overlap between left and right eigenstates of the Liouvillean, $I_i =  \la \bar{\rho}^{(i)} | \rho (0) \ra$.
We explicitly show how to find this conserved quantities in
the Supplement.
The conserved quantities can directly be used to calculate the
stationary value of observables for any initial state.
The conserved quantities {\it encode} certain information about the initial state into
the stationary state.
This is demonstrated in Fig.~\ref{fig:diss}.

We also investigated the question of the robustness of the SUSY-like
dynamics when we are detuned from the SUSY line.
We observed that the additional integral of motion, $I_{2}$, becomes a
time-dependent function with an extremely slow decay.
Namely, for deviations up to  $\delta g_{1,2}/\omega \sim
0.1$ from the SUSY line, the decay can be fitted with an exponential
function $I_{2}(t) \sim \exp(-\kappa t)$ with $\kappa\sim 10^{-3}$ for a very long time intervals corresponding to the scale of Fig.~\ref{fig:diss}.
This demonstrate a robustness of the SUSY-related dynamical properties even outside of the SUSY line.  
From a more general viewpoint this brings an analogy with the
classical KAM theory, where the invariant tori stay stable for a long
time.

{\it Cross-links: dynamics}
Time evolution starting from a given initial state is very natural for the quantum optics. In JC model when evolution starts with a coherent state one observes Rabi oscillations with a frequency $\bar{\Omega}_{R}=\sqrt{g^{2}_{R}\bar{n}+\Delta^{2}/4}$, their collapse and revival for average number of photons $\bar{n}>>1$ \cite{ENS-M},\cite{RWK}. In general, three time scales can be identified: Rabi oscillation period $(2\bar{\Omega})^{-1}$, their collapse time $T_{c}=\bar{\Omega}/g^{2}\sqrt{\bar{n}}$ and the revival time $T_{r}=2\pi\bar{\Omega}/g^{2}$. What would be an interpretation of these phenomena in terms of the Rashba model?
Consider the operator $\bar{\rho}_{q}=\exp(-|q|^{2}/4)\tau_{q}$, where $\tau_{q}=\exp[(-iqa-iq^{*}a^{\dag})/\sqrt{2}]$ is a displacement operator then. This operator is nothing but a generator of the GMP algebra of lowest Landau Level projected density operators \cite{GPM}, satisfying $[\tau_{q},\tau_{p}]=2i\tau_{q+p}\sin(\frac{q\wedge p}{2})$, where $q\wedge p=l^{2}({\bf q}\times{\bf p})\cdot\hat{\bf z}$ and $l=(\hbar c/eB_{0})^{1/2}$. Therefore, by preparing the condensed matter Rashba system in the eigenstate $|q\rangle$ of the projected density operator $\bar{\rho}_{q}|q\rangle=(iq\sqrt{2})|q\rangle$ one should be able to observe collapse and revivals of the Rabi oscillations.

Still another example of cross-links between quantum optical models and spin-orbit coupled condensed matte systems could be provided by the Ramsey $\pi$-pulse scheme (kicks) applied to the two-level subsystem \cite{Morigi}. Following the previous analogy with JC model one can suggest a Ramsey spectrometry magnetic field pulse scheme to measure decoherence effects in the Rashba model.

\begin{figure}[h!]
  \begin{center}
      \includegraphics[scale=1.0]{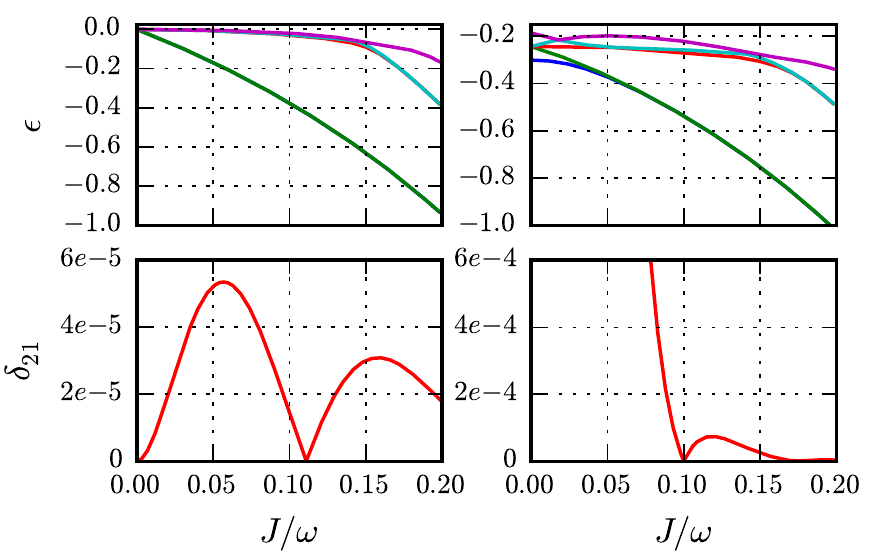}
    \caption{Top panels: The spectrum of a one dimensional array of 3 coupled
      resonators, each described by the generalized Rabi model, as a
      function of the hopping amplitude $J$ between the resonators for
      $\omega=1$ and $\Delta=2$.
      Bottom panels: Energy difference of the lowest two levels
      $\delta_{21}=E_{2}-E_{1}$.
      On the left panels the parameters are such that each generalized Rabi
      cavity is on the SUSY line, $g_{1}=1.5$ and $g_{2}=0.5$. On the
      right panels the parameters are chosen not to satisfy the SUSY
      condition, $g_{1}=1.4$ and $g_{2}=0.5$.
      } 
    \label{lattice}
  \end{center}
\end{figure}

{\it Coupled systems: prospects for quantum simulation of the SUSY field theories.}
We coupled several (up to three) cavities tuned to the SUSY line and observed persistent degeneracy of the ground state in a range of tunneling parameter, see Fig.~(\ref{lattice}).  A number of recent studies suggest that JC- or Rabi-coupled systems undergo the Mott insulator-superfluid transition, and e.g. in the weak tunneling limit the coupled systems can be mapped to the effective XY-model in magnetic field (similarly to \cite{tureci2}). At the SUSY point to include the effect of the tunneling term one should use a degenerate perturbation theory. This leads to the XY-model without effective magnetic field. Starting from two cavities and transforming to the bonding unti-bonding basis it is easy to show that the doubly degenerate SUSY line will exist in parameter space, although its position is altered by the tunneling rate. We conjecture that in the continuum limit coupled gR cavities could be described by the continuum SUSY field theory at specific parameter manifold. We do not exclude that the continuum model could have the critical line in parameter space where the effective theory is a super-conformal field theory. This issue will be addressed elsewhere. Another possibility to observe SUSY would be to design a system which is described by $H=\int dx Q^{\dag}(x)Q(x)$, where $Q(x)$ is a continuum analogue of $Q$ introduced here.

{\it Discussion}.
Further connection between dissipative dynamics of the optical model and spin-orbit coupled system can be foreseen in view of the finding of \cite{BOZ}: in zero magnetic field when $g_{1}=g_{2}$ there is an $SU(2)$ dynamical symmetry leading to non-diffusive spin transport in disordered spin-orbit coupled system. The SUSY we found here has the same effect on transport for $g_{1}^{2}- g_{2}^{2}=\Delta\omega$ and nonzero field $\Delta$.  

In \cite{Ciuti} it was found that the parity operation of the model consist of an electric and magnetic discrete transformations. These transformations are best defined in terms of the electric and magnetic coupling constants $\Omega_{E,B}=g_{1}\pm g_{2}$ respectively. By breaking these symmetries separately in the generalized version of the Dicke model establishes separate electric and magnetic phases. It is interesting to note that in this picture our SUSY line (\ref{susy-line}) is $\Omega_{E}\Omega_{B}=\omega\Delta$ and corresponds to the electro-magnetic self-dual line in the parameter space, invariant under exchange $\Omega_{E}\leftrightarrow\Omega_{B}$.

Our observation of nontrivial structure of the stationary state density matrix forming $SU(2)$ manifold suggests to think about nontrivial topology of the density matrix encoded in dissipative dynamics and possible classification of topologically non-equivalent stationary state density matrices. Initial state density matrix is mapped to the stationary state subspace. This implies that the initial state information will be partially stored in the compact space of stationary state manifold. This should be useful for the (partial) decoherence-free algorithms in quantum information.

{\it Acknowledgement} 
MT and VG acknowledge support of Swiss NSF and Delta Institute of Theoretical Physics (DITP); MT is also supported by the ECOST-STSM grant MP1210, Delta Institute for Theoretical Physics and the International Institute of Physics in Natal; MP is supported by the DFG.

\newpage
\clearpage
\onecolumngrid

\renewcommand{\theequation}{S\arabic{equation}} 
\renewcommand{\thepage}{S\arabic{page}} 
\renewcommand{\thesection}{S\arabic{section}}  
\renewcommand{\thetable}{S\arabic{table}}  
\renewcommand{\thefigure}{S\arabic{figure}}
\renewcommand{\bibnumfmt}[1]{[{\normalfont S#1}]}
\setcounter{page}{0}
\setcounter{equation}{0}

\begin{center}
{\bf \Large Supplementary Material}
\end{center}
\section{Possible physical realizations of the generalized Rabi model}
Here we overview several physical realizations of the generalized Rabi model. These examples include: (i) the model of spin-orbit interacting two-dimensional electron gas in external magnetic field; (ii) electric-magnetic coupling of light and matter and (iii) effective realization of the model using 3- and 4-level emitters.

\subsection{Rashba-Dresselhaus model in a magnetic field}
For the perpendicular magnetic field ${\bf B}=B_{0}{\bf e}_{z}$ the Hamiltonian we consider is
\beq
H_{RD}&=&\frac{\Pi^{2}_{x}+\Pi^{2}_{y}}{2m^{*}}+g^{*}\mu_{B}B_{0}\sigma_{z}\\
&+&\frac{2\alpha_{R}}{\hbar}(\Pi_{x}\sigma_{y}-\Pi_{y}\sigma_{x})+\frac{2\alpha_{D}}{\hbar}(\Pi_{x}\sigma_{x}+\Pi_{y}\sigma_{y})\nonumber
\label{RD}
\eeq
here $\Pi_{x}=p_{x}-\frac{eB_{0}}{2c}y$, $\Pi_{y}=p_{y}+\frac{eB_{0}}{2c}x$ are momentum operators in symmetric gauge, $\alpha_{R}$ is a Rashba coupling, $\alpha_{D}$ is a Dresselhaus coupling, $m^{*}$ is an effective mass, $g^{*}$ is a gyromagnetic ratio and $\mu_{B}=e\hbar/2mc$ is the Bohr magneton. 

Due to the commutation relation between momentum operators , $[\Pi_{x},\Pi_{y}]=ie\hbar B_{0}/c$ one can introduce canonically conjugated operators $Q=-\Pi_{y}\sqrt{c/eB_{0}}$ and $P=-\Pi_{x}\sqrt{c/eB_{0}}$ satisfying $[Q,P]=i\hbar$. Further, introducing $a=(Q+iP)/\sqrt{2\hbar}$ and $a^{\dag}=(Q-iP)/\sqrt{2\hbar}$ such that $[a,a^{\dag}]=1$ 
and making use of the $U(1)$ gauge (canonical) transformation $a\rightarrow ae^{i\pi/4}$, $\sigma_{+}\rightarrow \sigma_{+}e^{-i\pi/4}$ we arrive at the  expression which has exactly the form of the combination of the Jaynes-Cummings and the Rabi models, namely having rotating and counter-rotating terms of different strength
\beq
H_{RD} &=&\hbar\omega a^{\dag}a+\frac{\Delta}{2}\sigma_{z}\nonumber\\
&+&g_{1}(a^{\dag}\sigma_{-}+a\sigma_{+})+g_{2}(a^{\dag}\sigma_{+}+a\sigma_{-})
\label{JC+R}
\eeq
where $\omega\equiv\omega_{c}=eB_{0}/m^{*}c$, 
$g_{1,2}=2\sqrt{2}\hbar^{3/2}\kappa_{R,D}\omega_{c}$, $\Delta=\hbar\gamma\omega_{c}$ and $\kappa_{R,D}=\alpha_{R,D}\sqrt{eB_{0}}/\hbar^{2}\sqrt{c}$ while $\gamma=g^{*}m^{*}/2m$ is a gyromagnetic ratio.

\subsection{Dipole-magnetic coupling}
A complete form of the coupling between the electromagnetic field and the two-level system includes the electric dipole and magnetic couplings
\beq
H&=&H_{\mbox{atom}}+H_{\mbox{field}}+H_{\mbox{int}}\\
H_{\mbox{atom}}&=&\frac{\Delta}{2}\sigma^{z}\\
H_{\mbox{field}}&=&\frac{1}{2}({\bf E}^{2}+{\bf B}^{2})\\
H_{\mbox{int}}&=&{\bf E}\cdot \hat{\bf d}+\hat{\bf\mu}\cdot {\bf B}.
\eeq
In the second quantized picture $E\sim(a+a^{\dag})$ while while $B\sim i(a-a^{\dag})$. Moreover, in terms of the Pauli matrices, 
\beq
\hat{\bf d}=\Omega_{E} \sigma_{x},\qquad
\hat{\bf \mu}=\Omega_{B}\sigma_{y}
\eeq
where $\Omega_{E}=g_{1}+g_{2}$ and $\Omega_{B}=g_{1}-g_{2}$
In Ref.~\cite{Ciuti} it was shown that the conserved parity symmetry of the generalized Dicke model is a composite action of
\beq
& &P_{E}: \qquad E\rightarrow - E,\qquad d\rightarrow - d; \qquad B\rightarrow B, \qquad \mu\rightarrow \mu\\
& &P_{B}: \qquad B\rightarrow - B,\qquad \mu\rightarrow -\mu; \qquad E\rightarrow E, \qquad d\rightarrow d
\eeq 
We note that the SUSY line $\Omega_{E}\Omega_{B}=\omega\Delta$ is invariant under electric-magnetic transformation $d\leftrightarrow\mu$.


\subsection{Effective three-level scheme}
Consider the three-level $\Lambda$-system defined by the following Hamiltonian in the rotating frame \cite{3-level}
\beq
H&=&e^{i\Delta_{1}t}\Omega_{1}X^{23}+e^{i\Delta_{2}t}\Omega_{2}X^{13}+h.c.\\
&+& (e^{i\Delta_{1}t}\tilde{g}_{1}X^{23}+e^{i\Delta_{2}t}\tilde{g}_{2}X^{13}) a+h.c.\\
&+& \frac{\omega_{0}}{2}(X^{22}-X^{11})
\eeq
where $\Omega_{1,2}$ are the Rabi frequencies between $b-e$ and $a-e$ respectively and $\Delta_{1,2}$ are detunings for $a-e$ and $b-e$ transitions. The transition between levels $a$ and $b$ is given by the Rabi frequency $\omega_{0}/2$ and the energy of the level $b$ is larger. The quantum field $a, a^{\dag}$ couples levels $a-e$ and $b-e$ with the strengths $g_{1,2}$. The operators $X^{\alpha\beta}=|\alpha\rangle\langle\beta|$ ($\alpha,\beta=a,b,e$ describe transitions between corresponding energy levels. 

Using adiabatic elimination of the level $e$ we can deduce the following effective hamiltonian for the levels $a,b$ coupled by the quantum radiation field $a$
\beq
H_{eff}&=&-\left((\frac{\tilde{g}_{1}^{2}}{\Delta_{1}}+\frac{|\Omega_{2}|^{2}}{\Delta_{2}}) X^{11}+(\frac{\tilde{g}_{2}^{2}}{\Delta_{2}} +\frac{|\Omega_{1}|^{2}}{\Delta_{1}})X^{22}\right)a^{\dag}a\\
&-& (g_{1} X^{12}+g_{2}X^{21})a+ h.c.\\
&+&\frac{\omega_{0}}{2}(X^{22}-X^{11})
\eeq
where $g_{1,2}=\tilde{g}_{1,2}\Omega^{*}_{1,2}/\Delta_{1,2}$. Therefore the effective couplings $\tilde{g}_{1,2}$ can be tuned in a wide range by changing the detunings and Rabi frequencies. When transition between $a$ and $b$ is not allowed, the frequency $\omega_{0}=0$. The effective Hamiltonian thus has a form of generalized Rabi model with effective Bloch-Siegert shift. If however we impose the condition $\frac{g_{1}^{2}}{\Delta_{1}}+\frac{|\Omega_{2}|^{2}}{\Delta_{2}}=\frac{g_{2}^{2}}{\Delta_{2}} +\frac{|\Omega_{1}|^{2}}{\Delta_{1}}$ we end up with the generalized Rabi model. 

Similar considerations applied to the four-level atomic scheme lead to the same effective model \cite{4-level}.

\section{Matrix representation of supercharges}
The following matrix representation of supercharge for the JC model has been suggested in \cite{AL}
\beq
Q=\left(
\begin{array}{cccc}
0  &  0 &  \alpha & \gamma a \\
0 &   0 & \beta a^{\dag}  & \delta\\
 \alpha^{*} &\beta^{*} a   & 0 & 0   \\
 \gamma^{*} a^{\dag} & \delta^{*} & 0 & 0
\end{array}
\right)
\label{JCsupercharge}
\eeq

They found that for the JC Hamiltonian at {\it zero} detuning and shifted energy levels $H_{c}=H+cI$ the two super partners exist,
\beq
H=\left(
\begin{array}{cc}
 H_{1} & 0      \\
 0 &   H_{2}   
\end{array}
\right)
\eeq
where  $H_{1}=\hbar\omega (a^{\dag}a+\sigma_{3})+g(a^{\dag}\sigma_{-}+a\sigma_{+})$, $H_{2}=\hbar\omega (a^{\dag}a+\sigma_{3})+ig(a^{\dag}\sigma_{-}-a\sigma_{+})$ and $c=\frac{1}{2}\hbar\omega+g^{2}/(4\hbar\omega)$. In that case the parameters in (\ref{JCsupercharge}) are given by
\beq
\alpha=\frac{g}{2\sqrt{\hbar\omega}},\qquad \beta=\sqrt{\hbar\omega},\qquad \gamma=-i\sqrt{\hbar\omega},\qquad \delta=\frac{-ig}{2\sqrt{\hbar\omega}}
\eeq
Then $H_{c}=Q^{2}$. 

Let us consider possible extensions of this observation. Consider the following ansatz
\beq
Q=\left(
\begin{array}{cccc}
0  &  0 &  \alpha & \gamma_{1} a+\gamma_{2}a^{\dag} \\
0 &   0 & \beta_{1} a^{\dag} +\beta_{2}a & \delta\\
 \alpha^{*} &\beta^{*}_{1} a  +\beta_{2}^{*}a^{\dag} & 0 & 0   \\
 \gamma^{*}_{1} a^{\dag} +\gamma_{2}^{*}a& \delta^{*} & 0 & 0
\end{array}
\right)
\label{Rsupercharge}
\eeq
and its square,
\beq
Q^{2}=\left(
\begin{array}{cccc}
 |\alpha|^{2}+ C_{1}& G_{1}  & 0 &0 \\
 G_{1}^{*}&  C_{2}+|\delta|^{2}&0   &0\\
 0 &0   &  |\alpha|^{2}+C_{3}& G_{2} \\
  0&0   &G_{2}^{*}.  &C_{4}+|\delta|^{2}
\end{array}
\right)
\eeq
where 
\beq
G_{1}&=&\alpha(\beta_{1}^{*}a+\beta_{2}^{*}a^{\dag})+\delta^{*}(\gamma_{1}a+\gamma_{2}a^{\dag})\\
G_{2}&=&\delta(\beta_{1}^{*}a+\beta_{2}^{*}a^{\dag})+\alpha^{*}(\gamma_{1}a+\gamma_{2}a^{\dag})\\
C_{1}&=&(\gamma_{1}a+\gamma_{2}a^{\dag})(\gamma_{1}^{*}a^{\dag}+\gamma_{2}^{*}a)\\
C_{2}&=&(\beta_{1}a+\beta_{2}a^{\dag})(\beta_{1}^{*}a+\beta_{2}^{*}a^{\dag})\\
C_{3}&=&(\beta_{1}^{*}a+\beta_{2}^{*}a^{\dag})(\beta_{1}a+\beta_{2}a^{\dag})\\
C_{4}&=&(\gamma_{1}^{*}a^{\dag}+\gamma_{2}^{*}a)(\gamma_{1}a+\gamma_{2}a^{\dag})
\eeq

We would like to have $C_{1-4}$ diagonal and therefore we apply Bogoliubov transformation
\beq
a=\mu A+\nu A^{\dag}, \qquad a^{\dag}=\mu^{*}A^{\dag}+\nu^{*} A,\qquad |\mu|^{2}-|\nu|^{2}=1
\eeq
Transformation applied to $C_{1}$:
\beq
C_{1}&\rightarrow& (A^{\dag})^{2}\left(\gamma_{1}^{*}\gamma_{2}(\mu^{*})^{2}+\gamma_{1}\gamma_{2}^{*}\nu^{2}+|\gamma_{1}|^{2}\nu\mu^{*}+|\gamma_{2}|^{2}\mu^{*}\nu\right)\\
&+&(A)^{2}\left(\gamma_{1}^{*}\gamma_{2}(\nu^{*})^{2}+\gamma_{1}\gamma_{2}^{*}\mu^{2}+|\gamma_{1}|^{2}\mu\nu^{*}+|\gamma_{2}|^{2}\nu^{*}\mu\right)\\
&+&A^{\dag}A\left(2\gamma_{1}^{*}\gamma_{2}\mu^{*}\nu^{*}+2\gamma_{1}\gamma_{2}^{*}\mu\nu+(|\gamma_{1}|^{2}+|\gamma_{2}|^{2})(|\nu|^{2}+|\mu|^{2})\right)\\
&+&I\left(\gamma_{1}^{*}\gamma_{2}\mu^{*}\nu^{*}+\gamma_{1}\gamma_{2}^{*}\mu\nu+|\gamma_{1}|^{2}|\mu|^{2}+|\gamma_{2}|^{2}|\nu|^{2}\right)
\eeq
For the $C_{2}$ expression all $\gamma_{1,2}$ are to be replaced by $\beta^{*}_{1,2}$.
Transformation for the $G_{1}$:
\beq
G_{1}&\rightarrow&  A\left(\alpha(\beta_{1}^{*}\mu+\beta_{2}^{*}\nu^{*})+\delta(\gamma_{1}\mu+\gamma_{2}\nu^{*})\right)\\
&+& A^{\dag}\left(\alpha(\beta_{1}^{*}\nu+\beta_{2}^{*}\mu^{*})+\delta(\gamma_{1}\nu+\gamma_{2}\mu^{*})\right)
\eeq

Vanishing of terms $A^{2}$ and $(A^{\dag})^{2}$ leads to (for the first block)
\beq
(\gamma_{1}\nu+\gamma_{2}\mu^{*})(\gamma_{2}^{*}\nu+\gamma_{1}^{*}\mu^{*})=0\\
(\gamma_{1}\mu+\gamma_{2}\nu^{*})(\gamma_{2}^{*}\mu+\gamma_{1}^{*}\nu^{*})=0\\
(\beta_{1}^{*}\mu+\beta^{*}_{2}\nu^{*})(\beta_{2}\mu+\beta_{1}\nu^{*})=0\\
(\beta_{1}^{*}\nu+\beta^{*}_{2}\mu^{*})(\beta_{2}\nu+\beta_{1}\mu^{*})=0.
\label{vanish-quadratic}
\eeq
On the other hand we would like to have both $A$ and $A^{\dag}$ in the off-diagonal terms,
\beq
\delta^{*}(\gamma_{1}\nu+\gamma_{2}\mu^{*})+\alpha(\beta_{1}^{*}\nu+\beta_{2}^{*}\mu^{*})\neq 0\\
\delta(\gamma_{1}^{*}\nu^{*}+\gamma_{2}^{*}\mu)+\alpha^{*}(\beta_{1}\nu^{*}+\beta_{2}\mu)\neq 0\\
\alpha(\beta_{1}^{*}\mu+\beta_{2}^{*}\nu^{*})+\delta^{*}(\gamma_{1}\mu+\gamma_{2}\nu^{*})\neq 0\\
\alpha^{*}(\beta_{1}\mu^{*}+\beta_{2}\nu)+\delta(\gamma_{1}^{*}\mu^{*}+\gamma_{2}^{*}\nu)\neq 0
\label{equal}
\eeq
and similarly when $\gamma_{1,2}$ replaced by $\beta^{*}_{1,2}$.

Coefficient in front of $A^{\dag}A$ in the expression for $C_{1}$ and $C_{2}$ must be the same. Therefore
\beq
\left(2\gamma_{1}^{*}\gamma_{2}\mu^{*}\nu^{*}+2\gamma_{1}\gamma_{2}^{*}\mu\nu+(|\gamma_{1}|^{2}+|\gamma_{2}|^{2})(|\nu|^{2}+|\mu|^{2})\right)\\
=\left(2\beta_{1}\beta_{2}^{*}\mu^{*}\nu^{*}+2\beta_{1}^{*}\beta_{2}\mu\nu+(|\beta_{1}|^{2}+|\beta_{2}|^{2})(|\nu|^{2}+|\mu|^{2})\right)
\label{non-equal}
\eeq
One can reconcile these conditions which leads us to two  different solutions: one which is valid for the RWA while the other goes beyond RWA. 

\subsection{RWA-type models}
First possibility is to take a supercharge in the following form
\beq
Q=\left(
\begin{array}{cccc}
0  &  0 &  \alpha & \gamma a \\
0 &   0 & \beta a^{\dag}  & \delta\\
 \alpha^{*} &\beta^{*} a   & 0 & 0   \\
 \gamma^{*} a^{\dag} & \delta^{*} & 0 & 0
\end{array}
\right)
\label{JCsupercharge2}
\eeq
Its square is 
\beq
Q^{2}=
\left(
\begin{array}{cccc}
 |\alpha|^{2}+|\gamma|^{2}+|\gamma|^{2} a^{\dag}a & ga  & 0 &0 \\
 g^{*}a^{\dag} &   |\beta|^{2}a^{\dag}a +|\delta|^{2}&0   &0\\
 0 &0   &  |\alpha|^{2}+|\beta|^{2}+|\beta|^{2}a^{\dag}a &\tilde{g}a\\
  0&0   &\tilde{g}^{*}a^{\dag}   &|\gamma|^{2}a^{\dag}a+|\delta|^{2}
\end{array}
\right)
\eeq
where $g=\alpha\beta^{*}+\gamma\delta^{*}$, $\tilde{g}=\alpha^{*}\gamma+\delta\beta^{*}$. In the most general case we can put that $|\gamma|^{2}=\omega+\lambda$ while $|\beta|^{2}=\omega-\lambda>0$ and $|\alpha|^{2}+|\gamma|^{2}=\Delta_{1}+c_{1}$, $|\delta|^{2}=c_{1}-\Delta_{1}=c_{2}-\Delta_{2}>0$ where $c_{1,2}$ are some constant to be determined. 
Then $|\alpha|^{2}+|\beta|^{2}=\Delta_{2}+c_{2}$.
So, $H=\mbox{diag}(H_{1},H_{2})$ where
\beq
H_{1}&=&\omega a^{\dag}a +\lambda a^{\dag}a\sigma_{3}+\Delta_{1}\sigma_{3}+ga\sigma_{+}+g^{*}a^{\dag}\sigma_{-}+c_{1}\\
H_{2}&=&\omega a^{\dag}a -\lambda a^{\dag}a\sigma_{3}+\Delta_{2}\sigma_{3}+\tilde{g}a\sigma_{+}+\tilde{g}^{*}a^{\dag}\sigma_{-}+c_{2}
\eeq
with a constraint coming from equating expressions for $|\alpha|^{2}$: $\Delta_{1}+c_{1}-(\omega+\lambda)=\Delta_{2}+c_{2}-(\omega-\lambda)$ and from two definitions of $|\delta|^{2}$: $c_{1}-\Delta_{1}=c_{2}-\Delta_{2}$. From these two it follows that $\Delta_{1}-\Delta_{2}=\lambda$.

Focusing more on the $\lambda=0$, we find that $\Delta_{1}=\Delta_{2}=\Delta$ and $c_{1}=c_{2}=c$. Together with $\gamma=\beta=\sqrt{\omega}$ we have $\alpha+\delta^{*}=g/\sqrt{\omega}$. Since $|\alpha|^{2}-|\delta|^{2}=2\Delta-\omega$ we have $\alpha-\delta^{*} =(2\Delta-\omega)\sqrt{\omega}/g$ so that $2\alpha=(g/\sqrt{\omega})+d\sqrt{\omega}/g$ and $2\delta=(g/\sqrt{\omega})-d\sqrt{\omega}/g$ where $d=2\Delta-\omega$ is a detuning.

\subsection{non-RWA type models}
Second possibility is to take the supercharge in the following form
\beq
Q=\left(
\begin{array}{cccc}
0  &  0 &  \alpha & \gamma a \\
0 &   0 & \beta a & \delta\\
 \alpha^{*} &\beta^{*}a^{\dag} & 0 & 0   \\
 \gamma^{*}a^{\dag} & \delta^{*} & 0 & 0
\end{array}
\right)
\label{Rsupercharge2}
\eeq
and obtain for its square:
\beq
Q^{2}=\left(
\begin{array}{cccc}
 |\alpha|^{2}+|\gamma|^{2}+|\gamma|^{2}a^{\dag}a& \alpha\beta^{*}a^{\dag}+\gamma\delta^{*} a & 0 &0 \\
 \alpha^{*}\beta a+\gamma^{*}\delta a^{\dag} & |\beta|^{2}a^{\dag}a+|\beta|^{2}+|\delta|^{2} &0   &0\\
 0 &0   &  |\alpha|^{2}+|\beta|^{2}a^{\dag} a& \alpha^{*}\gamma a+ \delta\beta^{*} a^{\dag}\\
  0&0   &\alpha\gamma^{*} a^{\dag}+ \delta^{*}\beta a  &|\gamma|^{2}a^{\dag}a+|\delta|^{2}
\end{array}
\right)
\eeq
one can look into two cases:

1. $|\gamma|^{2}=\omega+\lambda$ while $|\beta|^{2}=\omega-\lambda$ such that $\omega>\lambda$. We demand that $|\alpha|^{2}+|\gamma|^{2}=c+\Delta$ while $|\beta|^{2}+|\delta|^{2}=c-\Delta$. Without lose of generality we assume that $\gamma$ and $\beta$ are real parameters. Then $\Delta=\lambda +(|\alpha|^{2}-|\delta|^{2})/2$, $c=\omega+(|\alpha|^{2}+|\delta|^{2})/2$ and $g_{1}=\delta^{*}\sqrt{\omega+\lambda}$, $g_{2}=\alpha\sqrt{\omega-\lambda}$ and $\tilde{g}_{1}=\alpha^{*}\sqrt{\omega+\lambda}$ and $\tilde{g}_{2}=\delta\sqrt{\omega-\lambda}$. The Hamiltonians $H_{1,2}$ are then 
\beq
H_{1}&=&\omega a^{\dag}a +\Delta\sigma_{z}+\lambda a^{\dag}a\sigma_{z}+g_{1}^{*}a^{\dag}\sigma_{-}+g_{1}a\sigma_{+}+g_{2}a^{\dag}\sigma_{+}+g_{2}^{*}a\sigma_{-}+cI\\
H_{2}&=&\omega a^{\dag}a +(\Delta-\lambda)\sigma_{z}-\lambda a^{\dag}a\sigma_{z}+\tilde{g}_{1}^{*}a^{\dag}\sigma_{-}+\tilde{g}_{1}a\sigma_{+}+\tilde{g}_{2}a^{\dag}\sigma_{+}+\tilde{g}_{2}^{*}a\sigma_{-}+(c-\omega)I\\
c&=&\omega+\frac{1}{2}\left(\frac{\omega(|g_{1}|^{2}+|g_{2}|^{2})+\lambda(|g_{2}|^{2}-|g_{1}|^{2})}{\omega^{2}-\lambda^{2}}\right)
\eeq

2. When $\lambda=0$ we have two sub cases: 

2a. $\Delta\neq 0$. In this case $|g_{2}|^{2}-|g_{1}|^{2}=2\Delta\omega$. Then $c_{1}=\omega+(|g_{2}|^{2}+|g_{1}|^{2})/2\omega$ for $H_{1}$ and $c_{2}=(|g_{2}|^{2}+|g_{1}|^{2})/2\omega$ for $H_{2}$.

2b. $\Delta=0$. Only in this case $g_{1}=g_{2}$ (Rabi model).

Note that we could also add other nonlinearities to the Hamiltonian, in particular the terms $\sigma a^{2}+\sigma^{*}a^{\dag}$. These terms could come from the quantization of the $A^{2}$ term (square of the vector potential which appears beyond the dipole approximation) in the Hamiltonian as well as from other radiative corrections. This term will not spoil dynamical SUSY and moreover it can be included into conserved SUSY. 

\subsection{Zero-mode Eigenfunctions}
First we determine conditions under which the supersymmetry is {\bf unbroken}. For this we need to satisfy a condition
\beq
Q|\Psi\rangle =0
\eeq
in the $4\times 4$ matrix representation of our $Q$-operators. 
Writing $|\Psi\rangle$ as 
\beq
\left(  \begin{array}{c}
    \psi_{1} \\ 
    \psi_{2} \\ 
    \psi_{3} \\
  \psi_{4} 
  \end{array}\right)
\eeq
we look into the possible solution as a series expansion in the Fock basis
\beq
|\psi_{j,n}\rangle=\sum_{n=0}^{\infty}c_{j,n}|n\rangle
\eeq
where coefficients $c_{j,n}$ ($j=1,2,3,4$) should satisfy orthonormality condition $\sum_{n}|c_{j,n}|^{2}=1$. 

Consider first the {\bf RWA case}. Applying $Q$ to our $|\Psi\rangle$. From the upper-right block we get two equations (remembering that $a|n\rangle=\sqrt{n}|n-1\rangle$, $a^{\dag}|n\rangle=\sqrt{n+1}|n+1\rangle$)
\beq
\sum_{n=0}^{\infty} \alpha c_{3,n}|n\rangle+\sum_{n=0}^{\infty}\gamma c_{4,n}\sqrt{n}|n-1\rangle =0\\
\sum_{n=0}^{\infty} \gamma c_{3,n}\sqrt{n+1}|n+1\rangle+\sum_{n=0}^{\infty}\delta c_{4,n}|n\rangle =0.
\eeq
Equating coefficients at the same state $|m\rangle$ we obtain a system
\beq
\alpha c_{3,n}+\gamma c_{4,n+1}\sqrt{n+1}=0\\
\gamma c_{3,n-1}\sqrt{n}+\delta c_{4,n}=0
\eeq
from which we get 
\beq
\frac{\alpha\delta}{\gamma^{2}}=n=\mbox{integer}
\eeq
Lower-left block gives the same condition with $\alpha,\delta$ replaced by their complex conjugates. When $\lambda=0$ we obtain $\alpha\delta/\gamma^{2}=[(g^{2}/\omega^{2})-(\delta^{2}/g^{2})]/4 =integer$.

Consider {\bf  non-RWA } case for $\lambda=0$. In this case one can convince yourself that the lower block can have only $|\psi_{1}=\psi_{2}=0$ as the only allowed solution. So we focus on the upper-right block. Similar reasoning like in the RWA case leads to the following set of equations 
\beq
\sum_{n=0}^{\infty} \alpha c_{3,n}|n\rangle+\sum_{n=0}^{\infty}\gamma c_{4,n}\sqrt{n}|n-1\rangle =0\\
\sum_{n=0}^{\infty} \gamma c_{3,n}\sqrt{n}|n-1\rangle+\sum_{n=0}^{\infty}\delta c_{4,n}|n\rangle =0.
\eeq
\beq
\alpha c_{3,n}+\gamma \sqrt{n+1} c_{4,n+1}=0\\
\gamma c_{3,n+1}\sqrt{n+1}+\delta c_{4,n}=0
\eeq
from which we obtain a nontrivial recurrence relations, in particular
\beq
\frac{c_{4,n+2}}{c_{4,n}}=\frac{\alpha\delta}{\gamma^{2}\sqrt{(n+1)(n+2)}},
\eeq
and the same for the $c_{3,n}$. In the case of recurrence relation which couples only nearest neighbor indexes one could determine the coefficients using normalization condition. Here we can express all coefficients in terms of $c_{4,0}$ and $c_{4,1}$. Denoting $z=\alpha\delta/\gamma^{2}$ we get (we suppress index 4 for a moment). 
\beq
c_{2}=\frac{z}{\sqrt{1\cdot 2}}c_{0},\quad c_{4}=\frac{z^{2}}{\sqrt{1\cdot2\cdot 3\cdot 4}}c_{0}\quad c_{6}=\frac{z^{3}}{\sqrt{1\cdot 2\cdot 3\cdot 4\cdot 5\cdot 6}}c_{0}, etc.\qquad c_{2n}=\frac{z^{n}}{\sqrt{(2n)!}}c_{0}\\
c_{3}=\frac{z}{\sqrt{2\cdot 3}}c_{1},\quad c_{5}=\frac{z^{2}}{\sqrt{1\cdot2\cdot 3\cdot 4 \cdot 5 }}c_{1}\quad c_{7}=\frac{z^{3}}{\sqrt{1\cdot2\cdot 3\cdot 4\cdot 5\cdot 6\cdot 7}}c_{1}, etc.\qquad c_{2n+1}=\frac{z^{n}}{\sqrt{(2n+1)!}}c_{1}
\eeq
One can rewrite then normalization condition for $|\Psi\rangle$,  as $|\psi_{3}|^{2}+|\psi_{4}|^{2}=1$ which leads to
\beq
(\frac{\gamma^{2}}{\alpha^{2}}|c_{4,1}|^{2}+|c_{4,0}|^{2})\cosh |z| +(\frac{\delta^{2}}{\gamma^{2}}|c_{4,0}|^{2}+|c_{4,1}|^{2} )\frac{\sinh |z|}{|z|}=1
\label{norm-cond}
\eeq

The components $\psi_{3,4}$ of a spinor $|\Psi\rangle$ can be written as $\frac{c_{0}}{2}[D(z)+ D(-z)]|0\rangle+\frac{c_{1}}{2}[D(z)-D(-z)]|0\rangle$, where $D(\alpha)$ is a displacement operator which creates a coherent state. 

Two zero eigenmodes are distinguished by the parity operator $P=\exp(i\pi \hat{N})$ which map $z\rightarrow -z$.

\section{Dissipation: Degenerate zero eigenvalues of a Liovillian}
Let us consider the following master equation for a density matrix
\be
i \frac{d}{d t} | \rho (t) \ra = L | \rho (t) \ra .
\label{m_eq}
\ee
We use Dirac-like notations to indicate that in Liovillian space a density matrix is represented by a vector.

The Liovillian $L$ is a non-Hermitian operator, therefore its eigenvalues are complex-valued. Moreover, in a physically meaningful model their imaginary parts must be smaller than zero (causality principle -- no exponentially growing solutions).

\subsection{Topological arguments}

We focus here on the case of zero dephasing.
First we note that the stationary matrix $|\rho\ra_{st}$ is annihilated by $Q$, $Q|\rho_{st}\ra_{st}=0$. The dimension of the ground state manifold is directly related to the dimension of the {\it cohomology} space ${\cal H}_{Q}$ of the SUSY $Q$-operator $\mbox{dim}{\cal H}_{Q}=\mbox{dim}(\mbox{ker}(Q)/\mbox{im}(Q)$, where $\mbox{ker}Q$ is the space of solutions $Q|\chi\rangle=0$, while $\mbox{im}(Q)$ is the  space of all states which can be written as $Q|\psi\rangle$ for some $|\psi\rangle$. This dimension in our case is equal 2. It was shown by Witten in \cite{Witten82} that dim${\cal H}$ is unchanged by one-parametric family of non-unitary transformations. In our context these non-unitary conjugations are generated by the dissipative evolution, $\exp(Lt)$ where $L$ is the Liouvillean. Therefore dim$(\rho_{st})=(\mbox{dim}({\cal H}))^{2}$. The stationary state is thus insensitive to the details of $L$.
In addition, the Witten index, being a topological invariant according to the Atiah-Singer index theorem, is also invariant under a large class of deformations. In our case  $W_{\mbox{ind}}=\mbox{dim}({\cal H})=2$ which justifies why $\rho_{st}$ is a two by two matrix. This, combined with a unit trace condition and hermiticity gives an equivalence with the $SU(2)$ group.

\subsection{General theory of degenerate Liouvillian}

Suppose that we have a peculiar situation when there is more than one zero eigenvalue (more precisely, $- i 0^+$). We look for the corresponding left and right eigenvectors of $L$
\beq
\la \bar{\rho}^{(i)} | L  &=& 0, \\
L | \rho^{(i)} \ra &=& 0 ,
\eeq
where $i=1, \ldots , r$, and $r$ is the degeneracy. Note that for a general non-Hermitian operator left and right eigenvalues are not related to each other by any operation (conjugation, transposition, etc.), and thus they are independent of each other, besides the condition
\be
\la \bar{\rho}^{(i)} | \rho^{(j)} \ra = \delta_{ij}.
\ee 
So these two eigenvalue problems have to be solved separately. This is explicitly reflected by an additional bar symbol in the left eigenvector.

Then, the projector onto zero eigenspace is given by
\be
P_0 = \sum_{i=1}^r | \rho^{(i)} \ra \la \bar{\rho}^{(i)} | .
\ee

Let us now write down a solution for time dynamics [Eq.~\eqref{m_eq}]
\be
| \rho (t) \ra = e^{- i L t} | \rho (0) \ra .
\ee
Representing the Liovillian  
\be
L = \sum_{s=r+1}^{n^2} \lambda_s P_s
\ee
by a sum over all non-zero eigenvalues times corresponding projectors, we find
\be
| \rho (t) \ra = \left[ 1+ \sum_s P_s \left( e^{- i \lambda_s t} -1 \right) \right] | \rho (0) \ra =  \left[ P_0 + \sum_s P_s e^{- i \lambda_s t} \right] | \rho (0) \ra ,
\ee
where we used $P_0 + \sum_s P_s =1$.

In the long time limit we obtain the stationary density matrix
\be
| \rho (t \to \infty) \ra \equiv | \rho_{st}  \ra = P_0 | \rho (0) \ra  = \sum_i | \rho^{(i)} \ra \la \bar{\rho}^{(i)} | \rho (0) \ra .
\label{statlimit}
\ee

Let us now consider the quantities $I_i =  \la \bar{\rho}^{(i)} | \rho (0) \ra$. In this form they appear to depend on initial conditions (however, in absence of degeneracy this dependence is gone, see below). Acting  on \eqref{m_eq} with $P_0$, we find
\be
\sum_i | \rho^{(i)} \ra \frac{d}{dt} \la  \bar{\rho}^{(i)} | \rho (t) \ra =0.
\ee
Since all $| \rho^{(i)} \ra$ are linearly independent, we find that $I_i (t) = \la  \bar{\rho}^{(i)} | \rho (t) \ra$  are conserved quantities, $I_i (t) = I_i$.

The values of (most of) $I_i$ are fixed by initial conditions. But there is one conserved quantity, namely $\mathrm{Tr} [\rho (t)] =1$, which is independent of initial conditions. This means that the basis in the degenerate zero subspace can be always chosen in such a way that one left eigenvector appears to be $\la  \bar{\rho}^{(tr)}| = (1, \ldots, 1, 0, \ldots, 0)$, where entries $1$ appear in positions of diagonal density matrix elements ($n$ times), and entries $0$ appear in  positions of nondiagonal elements ($n^2 -n$ times). Thus, $\la  \bar{\rho}^{(tr)} | \rho (t) \ra = \mathrm{Tr} [\rho (t)] =1$.

In absence of degeneracy, $\la  \bar{\rho}^{(tr)}|$ is the only (left) eigenvector, and therefore the stationary density matrix reads
\be
| \rho_{st}  \ra = | \rho^{(tr)} \ra , 
\ee
i.e. it coincides with the corresponding right eigenvector. Thereby, all information about initial conditions is lost, as $| \rho^{(tr)} \ra$ depends only on $L$.

The knowledge of $\rho_{st}$ allows us to find expectation values of observables in the stationary regime.

\section{Relaxation in generalized Rabi model}

Define
\beq
\bar{\rho}^{(i'j')} = \sum_{k,k'} \bar{\rho}^{(i'j')}_{k k'} | k \rangle \langle k' | 
\eeq
which is equivalent to 
\beq
| \bar{\rho}^{(i'j')} \ra = \sum_{k,k'} \bar{\rho}^{(i'j')}_{k k'} | k   k' \ra
, \quad   \la \bar{\rho}^{(i'j')} |  = \sum_{k,k'} \bar{\rho}^{(i'j')*}_{k k'} \la k k' | =  \sum_{k,k'} (\bar{\rho}^{(i'j')\dagger})_{k' k} \la k   k' |
\eeq
where
\be
|k \rangle \langle k' | = | k k' \ra = \la k'k | .
\ee

It is apparent that $\la l l' |  k k' \ra = \mathrm{Tr} [ |l' \rangle \langle l |k \rangle \langle k' |] = \delta_{l k} \delta_{l' k'} $, and therefore
\be
\la \bar{\rho}^{(i'j')} | \rho^{(ij)} \ra =  \sum_{l,l',k,k'} (\bar{\rho}^{(i'j')\dagger})_{l' l}  \rho^{(ij)}_{k k'} \la l   l' | k   k' \ra = \mathrm{Tr} [\bar{\rho}^{(i'j')\dagger} \rho^{(ij) }].
\ee

First, we determine the subspace of stationary density matrices solving the equation
\beq
0 &=& - i [H, \rho^{(ij)} ] + \mathcal{D} \left[\sum_k \Phi_k |k \rangle \langle k | \right] \rho^{(ij)} + \sum_{k'>k} \mathcal{D} \left[ O^{(kk')} \right] \rho^{(ij)},
\label{stdm}
\eeq
where $O^{(kk')} = \sqrt{\Gamma^{(k k')}} | k \rangle \langle k' |$, and
\beq
\mathcal{D} [O] \rho &=& O \rho O^{\dagger} - \frac12 \rho O^{\dagger} O - \frac12  O^{\dagger} O \rho , \\
H &=& \sum_k \varepsilon_k  |k \rangle \langle k | . 
\eeq
Note that $\Gamma^{(kk')} \neq 0$ only for $k'>k$.

Suppose now that $\varepsilon_1 = \varepsilon_2$, and $\Gamma^{(12)} =0$. We find that: 1) there are four solutions $\rho^{(ij)} = | i \rangle \langle j|$, $i,j=1,2$, to the equation \eq{stdm} for $\Phi_1 = \Phi_2$; 2) there are two solutions $\rho^{(11)} = |1 \rangle \langle 1|$ and $\rho^{(22)} =  |2 \rangle \langle 2|$ for $\Phi_1 \neq \Phi_2$.

We establish the conserved quantities $\bar{\rho}^{(ij)} = \sum_{kk'} \bar{\rho}^{(ij)}_{kk'} | k \rangle \langle k'|$ solving the equations
\beq
\la \bar{\rho}^{(i'j')} | \rho^{(ij)} \ra = \delta_{ii'} \delta_{jj'}
\label{bi-orth}
\eeq
and
\beq
0 &=&  i [H, \bar{\rho}^{(ij)} ] +  \mathcal{D}^{\dagger} \left[\sum_k \Phi_k |k \rangle \langle k | \right] \bar{\rho}^{(ij)}+ \sum_{k'>k} \mathcal{D}^{\dagger} \left[ O^{(kk')} \right] \bar{\rho}^{(ij)},
\label{stcons}
\eeq
where
\beq
\mathcal{D}^{\dagger} [O] \rho &=& O^{\dagger} \rho O - \frac12 \rho O^{\dagger} O - \frac12  O^{\dagger} O \rho .
\eeq

The condition \eq{bi-orth} implies
\be
 \bar{\rho}^{(ij)}_{kk'} = \delta_{i k} \delta_{j k'}, \quad i,j=1,2; \quad k,k'=1,2 .
\ee
The other components $k,k'$ of $\bar{\rho}^{(ij)}$ should be found from the equation \eq{stcons}.

It is easy to check that $\bar{\rho}^{(12)} = |1 \rangle \langle 2|$ and  $\bar{\rho}^{(21)} = |2 \rangle \langle 1|$ for $\Phi_1 = \Phi_2$, and for $\Phi_1 \neq \Phi_2$ these states are of no interest. Therefore we can concentrate on the diagonal components $\bar{\rho}^{(ii)} \equiv \bar{\rho}^{(i)}$, $i=1,2$, such that $\bar{\rho}^{(1)}_{11}= \bar{\rho}^{(2)}_{22}=1$ and $\bar{\rho}^{(1)}_{22}= \bar{\rho}^{(2)}_{11}=0$. We rewrite the equation \eq{stcons} for $\bar{\rho}^{(i)}_{kk'} $ as
\beq
0 &=& i \sum_{k=3}^{\infty} \sum_{k'=3}^{\infty} (\varepsilon_k - \varepsilon_{k'}) \bar{\rho}^{(i)}_{kk'} |k \rangle \langle k'| \nonumber \\
&+& \sum_{k=3}^{\infty} \Gamma^{(i k)} |k \rangle \langle k|  -  \sum_{k=3}^{\infty} \sum_{k'=3}^{\infty} \frac{\Gamma^{(1k)}+ \Gamma^{(1k')}+\Gamma^{(2k)}+ \Gamma^{(2k')}}{2} \bar{\rho}_{kk'}^{(i)} |k \rangle \langle k'| \nonumber \\
&+& \sum_{k=4}^{\infty} \left( \sum_{l=3}^{k-1} \Gamma^{(lk)} \bar{\rho}_{ll}^{(i)} \right) |k \rangle \langle k | -  \sum_{k=4}^{\infty}  \left( \sum_{l=3}^{k-1}  \frac{\Gamma^{(lk)}}{2} \right) \sum_{k'=1}^{\infty} \bar{\rho}^{(i)}_{kk'} |k \rangle \langle k' |  -  \sum_{k'=4}^{\infty}  \left( \sum_{l=3}^{k'-1}  \frac{\Gamma^{(lk')}}{2} \right) \sum_{k=1}^{\infty} \bar{\rho}^{(i)}_{kk'} |k \rangle \langle k' | .
\eeq
Note that the term with $\Phi_k$ drops out.
We immediately see that $\bar{\rho}^{(i)}_{kk'} =0$ for $k \neq k'$, because the corresponding equation is homogeneous. An equation for the diagonal components $\bar{\rho}^{(i)}_{kk} \equiv \bar{\rho}_k^{(i)}$ simplifies to
\beq
0 &=& \sum_{k=3}^{\infty} \Gamma^{(i k)} |k \rangle \langle k|  -  \sum_{k=3}^{\infty} (\Gamma^{(1k)}+\Gamma^{(2k)}) \bar{\rho}_{k}^{(i)} |k \rangle \langle k| \nonumber \\
&+& \sum_{k=4}^{\infty} \left( \sum_{l=3}^{k-1} \Gamma^{(lk)} \bar{\rho}_{l}^{(i)} \right) |k \rangle \langle k | -  \sum_{k=4}^{\infty}  \left( \sum_{l=3}^{k-1} \Gamma^{(lk)} \right)  \bar{\rho}^{(i)}_{k} |k \rangle \langle k | \nonumber \\
&=& \Gamma^{(i 3)} |3 \rangle \langle 3|  -  (\Gamma^{(13)}+\Gamma^{(23)}) \bar{\rho}_{3}^{(i)} |3 \rangle \langle 3| \nonumber \\
&+&  \sum_{k=4}^{\infty} \left( \Gamma^{(ik)} + \sum_{l=3}^{k-1} \Gamma^{(lk)} \bar{\rho}_{l}^{(i)} \right) |k \rangle \langle k | -  \sum_{k=4}^{\infty}  \left( \sum_{l=1}^{k-1} \Gamma^{(lk)} \right)  \bar{\rho}^{(i)}_{k} |k \rangle \langle k |, 
\eeq
from which we conclude
\beq
\bar{\rho}_{3}^{(i)} &=& \frac{\Gamma^{(i 3)}}{\Gamma^{(13)}+\Gamma^{(23)}}, \\
\bar{\rho}_{k}^{(i)} &=& \frac{\Gamma^{(ik)} + \sum_{l=3}^{k-1} \Gamma^{(lk)} \bar{\rho}_{l}^{(i)}}{\sum_{l=1}^{k-1} \Gamma^{(lk)}}.
\eeq
The latter expression is the recurrence relation, which allows to evaluate $\bar{\rho}_{k}^{(i)}$, if all $\bar{\rho}_{l<k}^{(i)}$ are known (recall that $\rho_1^{(1)}= \rho_2^{(2)} =1$ and $\rho_2^{(1)}= \rho_1^{(2)} =0$).

Note that $\bar{\rho}^{(1)}+ \bar{\rho}^{(2)} = \bar{\rho}^{(tr)} =1$: for $k=1,2,3$ this is obvious, while for $k \geq 4$ we obtain
\beq
\bar{\rho}_{k}^{(tr)} &=& \frac{\Gamma^{(1k)}+\Gamma^{(2k)}  + \sum_{l=3}^{k-1} \Gamma^{(lk)} \bar{\rho}_{l}^{(tr)}}{\sum_{l=1}^{k-1} \Gamma^{(lk)}}.
\eeq
If $\bar{\rho}_3^{(tr)}=1$, then $\bar{\rho}_4^{(tr)}=1$, and so forth, which leads to $\bar{\rho}_k^{(tr)} =1$ for all $k$.

The matrix $\bar{\rho}^{(tr)}$ is conjugated to the stationary matrix $\rho_{st} = \frac12 (|1 \rangle \langle 1| + |2 \rangle \langle 2 |)$. Another pair of conjugated matrices is $\bar{\rho}^{(diff)} = \bar{\rho}^{(1)} - \bar{\rho}^{(2)}$ and $\rho_{diff} =  \frac12 (|1 \rangle \langle 1| - |2 \rangle \langle 2 |)$.

According to \eq{statlimit} we find
\beq
\rho (t \to \infty) &=& \frac12 (|1 \rangle \langle 1 | +|2 \rangle \langle 2 | )
+  \frac12 (|1 \rangle \langle 1 | -|2 \rangle \langle 2 | ) \mathrm{Tr} \left[ (\bar{\rho}^{(1)} - \bar{\rho}^{(2)}) \rho (0)\right] \\
&+& |1 \rangle \langle 1 | \rho (0) | 2 \rangle \langle 2 | +  |2 \rangle \langle 2 | \rho (0) | 1 \rangle \langle 1 |, 
\eeq 
where the second line added for the case $\Phi_1 = \Phi_2$ (e.g., when both of them are zero).


\begin{thebibliography}{99}
 
\bibitem{weinberg}
S. Weinberg, {\it The Quantum Theory of Fields} (Cambridge University Press, 2005).

\bibitem{susy-cosmology}
J. L. Feng, Annals of Physics {\bf 315}, 2 (2005).

\bibitem{spintronics}
I. Zuti\'{c}, J. Fabian, S. Das Sarma, Rev. Mod. Phys. {\bf 76}, 323 (2004).

\bibitem{top-ins}
M. Z. Hasan, C. L. Kane, Rev. Mod. Phys. {\bf 82}, 3045 (2010);
X.-L. Qi, S.-C. Zhang, {\it ibid} {\bf 83}, 1057 (2011).

\bibitem{majorana}
V. Mourik, et al. Science {\bf 336}, 1003 (2012).

\bibitem{so-cold-gases}
Y.-J. Lin, et al. Phys. Rev. Lett. {\bf 102}, 130401 (2009); Y.-J. Lin, et al. 
Nature {\bf 462}, 628 (2009); Y.-J. Lin, et al. 
Nature Phys. {\bf 7}, 531 (2011); Y.-J. Lin, K. Jimenez-Garcia, and I. B. Spielman, Nature {\bf 471}, 83 (2011).
R. A. Williams, et al. 
Science {\bf 335}, 314 (2012).
L. W. Cheuk, et al. 
Phys. Rev. Lett. {\bf 109},
095302 (2012).


\bibitem{cav-QED-1}
S. Haroche and J. M. Raimond, {\it Exploring the Quantum: Atoms, Caviries and photons} (Oxford, Oxford University Press, 2006)

\bibitem{socreview}
V. Galitski and I. B. Spielman, Nature {\bf 494}, 49 (2013).


\bibitem{JC}
E. T. Jaynes, F. W. Cummings, Proc. Inst. Elect. Eng. {\bf 51}, 89 (1963); F. W. Cummings, Phys. Rev. {\bf 140}, A1051 (1965). 


\bibitem{Rabi}
I. I. Rabi, Phys. Rev. {\bf 49}, 324 (1936); {\bf 51}, 652 (1937).



\bibitem{braak}
D. Braak, Phys. Rev. Lett. {\bf 107}, 100401 (2011).


\bibitem{irish}
E. K. Irish, J. Gea-Banacloche, I. Martin, and K. C. Schwab, Phys. Rev. B {\bf 72}, 195410 (2005).



\bibitem{circ-QED-2}
D. I. Schuster, {\it et al.} Nature {\bf 445} 515 (2007); M. Hofheinz, {\it et al.}  Nature {\bf 459} 546 (2009).

\bibitem{circ-QED-3}
P. Forn-Diaz, {\it et al.} Phys. Rev. Lett. {\bf 105} 237001 (2010); T. Niemczyk {\it et al.}, Nature Phys. {\bf 6},  772 (2010).


\bibitem{3-level}
J. Cho, D. G. Angelakis, S. Bose, Phys. Rev. A {\bf 78}, 062338 (2008).

\bibitem{4-level}
F. Dimer, B. Estienne, A. S. Parkins, and H. J. Carmichael, Phys. Rev. A {\bf 75}, 013804 (2007).


\bibitem{grimsmo4level}
A. L. Grimsmo and S. Parkins, Phys. Rev. A 87, 033814 (2013).



\bibitem{Junker}
G. Junker, {\it Supersymmetric Methods in Quantum and Statistical Physics}, Springer (1996).

\bibitem{CKS}
F. Cooper, A. Khare, U. Sukhatme, {\it Supersymmetry in Quantum Mechanics}, World Scientific (2001).


\bibitem{atiyah}
M. F. Atiyah and I. M. Singer, Annals of Mathematics 87, pp. 484 (1968).
M. F. Atiyah and I. M. Singer, Annals of Mathematics 87, pp. 546 (1968).



\bibitem{LR}
H. Lewis, W. B. Riesenfeld, J. Math. Phys. {\bf 10}, 1458 (1969).

\bibitem{Nicolai}
H. Nicolai, J. Phys. A: Math. Gen. {\bf 9}, 1497 (1976); {\it ibid} {\bf 10}, 2143 (1977).

\bibitem{Witten-susy}
E. Witten, Nucl. Phys. B {\bf 188}, 513 (1981). 

\bibitem{Witten}
E. Witten, J. Diff. Geom. {\bf 17}, 661 (1982). 

\bibitem{Witten82}
E. Witten, Nucl. Phys. B {\bf 202}, 253 (1982). 

\bibitem{susy-book}
L. Frappat, A. Sciarrino, P. Sorba, {\it Dictionary on Lie algebras and superalgebras},
Academic Press (2000); arXiv:hep-th/9607161.



\bibitem{tureci}
M. Schiro, M. Bordyuh, B. Oztop, H. E. Tureci,
Phys. Rev. Lett. {\bf 109}, 053601 (2012).

\bibitem{tureci2}
M. Schir\'{o}, M. Bordyuh, B. \"{O}ztop, H. E. T\"{u}reci, J. Phys. B: At. Mol. Opt. Phys. {\bf 46}, 224021(2013).


\bibitem{BOZ}
B. A. Bernevig, J. Orenstein, S.-C. Zhang, Phys. Rev. Lett. {\bf 97}, 236601 (2006).


\bibitem{AL}
V. A. Andreev, P. B. Lerner, Phys. Lett. A {\bf 134}, 507 (1989). 

\bibitem{RWK}
G. Rempe, H. Walther, N. Klein, Phys. Rev. Lett. {\bf 58}, 353 (1987).


\bibitem{ENS-M}
J. H. Eberly, N. B. Narozhny, J. J. Sanchez-Mondragon, Phys. Rev. Lett. {\bf 44}, 1323 (1980).


\bibitem{GPM}
S. M. Girvin, A. H. MacDonald, P. M. Platzman, Phys. Rev. B {\bf 33}, 2481 (1986). 

\bibitem{SNR}
M. Scheunert, W. Nahm, V. Rittenberg, J. Math. Phys. {\bf 18}, 155 (1977).

\bibitem{JG}
P. D. Jarvis, H. S. Green, J. Math. Phys. {\bf 20}, 2115 (1979). 


\bibitem{Morigi}
G. Morigi, E. Solano, B.-G. Englert, H. Walther, Phys. Rev. A {\bf 65}, 040102(R) (2002);
A. Auffeves, et al. 
Phys. Rev. Lett. {\bf 91}, 230405 (2003). 


\bibitem{CW}
H. J. Carmichael, and D. F. Walls, J. Phys. A: Math. Gen. {\bf 6}, 1552 (1973).

\bibitem{BP}
H.-P. Breuer, F. Petruccione, {\it The Theory of Open Quantum Systems} (Oxford UP, 2007).

\bibitem{BGB}
F. Beaudoin, J. M. Gambetta, and A. Blais, Phys. Rev. A {\bf 84}, 043832 (2011).

\bibitem{AHE}
S. Agarwal, S. M. Hashemi Rafsanjani, J. H. Eberly, arXiv:1304.5308.

\bibitem{Ciuti}
A. Baksic, C. Ciuti, arxiv1310.3780.


\end{thebibliography}
\end{document}